\documentclass[structabstract]{aa}  
\usepackage{amsmath}
 \usepackage{natbib}                                   % (traditional abstract) 
 \usepackage{upgreek}
\usepackage{graphicx}
\usepackage{txfonts}
\usepackage{color}
\begin{document}
   \title{A {\it Herschel } [C\,{\sc ii}] Galactic plane survey III:  [C\,{\sc ii}] as a tracer of star formation}

   \author{J. L. Pineda,  W. D. Langer,  and P. F. Goldsmith
          }

\authorrunning{Pineda, Langer,  and Goldsmith}
\titlerunning{[C\,{\sc ii}] as a tracer of star formation}

   \institute{Jet Propulsion Laboratory, California Institute of Technology, 4800 Oak Grove Drive, Pasadena, CA 91109-8099, USA\\
              \email{Jorge.Pineda@jpl.nasa.gov}
             }

   \date{Received 25 April 2014/ Accepted 27 August 2014}

% \abstract{}{}{}{}{} mediu
% 5 {} token are mandatory
 
  \abstract{ The [C\,{\sc ii}] 158$\mu$m line is the brightest
far--infrared cooling line in galaxies, representing 0.1 to 1\% of
their FIR continuum emission, and is therefore a potentially powerful
tracer of star formation activity. The [C\,{\sc ii}] line traces
different phases of the interstellar medium (ISM), including the
diffuse ionized medium, warm and cold atomic clouds, clouds in
transition from atomic to molecular, and dense and warm photon
dominated regions (PDRs). Therefore without being able to separate the
contributions to the [C\,{\sc ii}] emission, the relationship of this
fine structure line emission to star formation has been unclear.
 } {We study the relationship between the [C\,{\sc ii}] emission and
the star formation rate (SFR) in the Galactic plane and separate the
relationship of different ISM phases to the SFR. We compare these
relationships to those in external galaxies and local clouds, allowing
examinations of these relationships over a wide range of physical
scales.  }
   { We compare the distribution of the [C\,{\sc ii}] emission, with
    its different contributing ISM phases, as a function of
    Galactocentric distance with the SFR derived from radio continuum
    observations.  We also compare the SFR with the surface density
    distribution of atomic and molecular gas, including the CO--dark
    H$_2$ component.} %
   { The [C\,{\sc ii}] and SFR are well correlated at Galactic scales
 with a relationship that is in general agreement with that found
 for external galaxies. By combining [C\,{\sc ii}] and SFR data points
 in the Galactic plane with those in external galaxies and nearby star
 forming regions, we find that a single scaling relationship between
 the [C\,{\sc ii}] luminosity and SFR applies over six orders of
 magnitude.  The [C\,{\sc ii}] emission from different ISM phases are
 each correlated with the SFR, but only the combined emission shows a
 slope that is consistent with extragalactic observations.
 These ISM components have roughly comparable contributions to the
 Galactic [C\,{\sc ii}] luminosity: dense PDRs (30\%), cold H\,{\sc i}
 (25\%), CO--dark H$_2$ (25\%), and ionized gas (20\%).
The SFR--gas surface density
 relationship shows a steeper slope compared to that observed in
 galaxies, but one that it is consistent with those seen in nearby
 clouds. The different slope is a result of the use of a constant
 CO--to--H$_2$ conversion factor in the extragalactic studies, which
 in turn is related to the assumption of constant metallicity in
 galaxies.  We find a linear correlation between the SFR surface
 density and that of the dense molecular gas.  } {} 
               
   \maketitle
%
%________________________________________________________________

\section{Introduction}

The [C\,{\sc ii}] 158\,$\mu$m line is the strongest far-infrared (FIR)
spectral line in galaxies, representing 0.1 to 1\% of their FIR
continuum emission \citep{Stacey1991,Genzel2000}.  The [C\,{\sc ii}]
line is an important ISM coolant and thus has been suggested to be a
powerful indicator of the star formation activity in galaxies
\citep[e.g.][]{Boselli2002,Stacey2010,deLooze2011,Sargsyan2012,Sargsyan2014,deLooze2014}.
Ionized carbon\footnote{Carbon is ionized by photons with energies
$\geq$11.6\,eV and C$^+$ can be excited by collisions with H, H$_2$
and electrons, with critical densities of $3\times10^3$\,cm$^{-3}$,
$6.1\times10^{3}$\,cm$^{-3}$ (at $T_{\rm kin}=$100\,K), and
44\,cm$^{-3}$ (at $T_{\rm kin}=$8000\,K ), respectively \citep[see
e.g.][]{Goldsmith2012}.} is present in a variety of phases of the
interstellar medium (ISM), including the diffuse ionized medium, warm
and cold atomic clouds, clouds making a transition from atomic to
molecular form, and dense and warm photon dominated regions
\citep[e.g.][]{Pineda2013}.  Since the [C\,{\sc ii}] line intensity is
very sensitive to the physical conditions of the line--emitting gas,
and as it be produced by gas that is not directly involved in the
process of star formation, the origin of its correlation with the star
formation activity is uncertain. This relationship can also be
affected by different environmental conditions and may vary depending
on the spatial scales at which it is studied.

In most extragalactic sources studied to date [C\,{\sc ii}] is
correlated with measurements of the star formation rate. However, in
some external galaxies, such as the ultra luminous infrared galaxies
(ULIRGs), observations suggest a deficit of [C\,{\sc ii}] emission
with respect to the far--infrared emission (used to trace the star
formation rate; e.g. \citealt{Malhotra1997,Luhman1998}).  This
deficit, however, applies only to the extreme environments of their
central regions, while their disks show a relationship between
[C\,{\sc ii}] and FIR emission that is similar to that observed in
normal galaxies \citep{DiazSantos2014}. Additionally,
ULIRGS represent a small fraction of the total population of galaxies
in the Universe, and therefore the potential error in the
determination of the star formation rate (SFR) from [C\,{\sc ii}] in
large samples of galaxies even including ULIRGs should be small.

The [C\,{\sc ii}] line is a potentially important tool for tracing
star formation activity as well as the physical conditions of distant
galaxies.  As it is a spectral line, it can be used to determine the
redshift and, if the line is resolved, to determine the dynamical mass
of galaxies. Additionally, as mentioned above, the [C\,{\sc ii}] line
is sensitive to the density and temperature of the line--emitting gas,
and therefore can be used as a diagnostic of physical conditions. For
these reasons, redshifted [C\,{\sc ii}] will be observed in a large
sample of galaxies with ALMA.  Studies at small spatial scales in the
local Universe are important for understanding how and why this line
traces star formation and how it can be used to determine the physical
conditions of the gas. Such studies are an important input for the
interpretation of the observations of [C\,{\sc ii}] in high--redshift
galaxies.

To understand the relationship between the [C\,{\sc ii}] emission
and star formation in external galaxies we need to establish their
relationship in our Galaxy.  However, our location within the Milky
Way requires that we use spectrally resolved [C\,{\sc ii}] to locate
the source of the emission as a function of radius (or position).  To
date the only velocity resolved Galactic survey of [C\,{\sc ii}] is
the Galactic Observations of C$^+$ (GOT C+) survey taken with HIFI
\citep{deGraauw2010} on {\it Herschel} \citep{Pilbratt2010}.  In
\citet{Pineda2013}, we used the GOT C+ survey to study the
distribution of the [C\,{\sc ii}] line as a function of Galactocentric
distance in the plane of the Milky Way.  We used the [C\,{\sc ii}]
emission, together with that of H\,{\sc i}, $^{12}$CO, and $^{13}$CO
to study the distribution of the different phases of the interstellar
medium in the plane of the Milky Way. In the present paper we compare
the distribution of the [C\,{\sc ii}] emission, as well as that of the
different phases of the ISM, with that of the star formation rate over
a wide range of physical conditions in the plane of the Milky Way.  By
studying the Galactic plane, we aim to provide a connection between
processes in individual star forming regions and those of entire
galaxies.

The determination of the star formation rate in the Milky Way is
complicated because the FUV or H$\alpha$ observations traditionally
used in external galaxies to measure the rate are highly extinguished
by the interstellar dust in the Galactic plane. Instead, several other
methods are used, including observations of radio free--free
(bremsstrahlung) emission \citep{Mezger1978,Guesten1982,Murray2010},
supernova remnants \citep{Reed2005}, pulsar counts \citep{Lyne1985},
and dust continuum emission \citep{Misiriotis2006}. When using the
same initial mass function (IMF) and models of massive stars, these
different methods agree that the total star formation rate of the
Galaxy is 1.9 M$_{\odot}$\,yr$^{-1}$ \citep{Chomiuk2011}. In this
paper we use radio continuum data to determine the radial distribution
of the number of Lyman continuum photons (proportional to the star
formation rate) that best reproduces the observed free--free emission
in the Galactic plane. We will then use the radial distribution of the
number of Lyman continuum photons to calculate the SFR as a function
of radius and compare it with the [C\,{\sc ii}] radial emission
profile.

The relationship between the star--formation activity and the gas
content in galaxies has been widely studied in recent decades. It has
been found that the star--formation rate per unit area and the gas
surface density in galaxies are strongly correlated
\citep{Schmidt1959,Kennicutt98}.  This correlation suggests that how
molecular clouds form out of atomic gas and how they evolve to form
dense cores, which are the sites where stars form, are fundamental
processes governing the evolution of galaxies.  \citet{Kennicutt98}
compared the star formation rate and gas (H\,{\sc i}+2H$_2$) surface
densities of a large sample of galaxies, deriving a scaling relation
characterized by a power law index of $1.4$. Later studies have shown
that the SFR is better correlated with the surface density of H$_2$
rather than that of H\,{\sc i}
\citep{Wong2002,Bigiel2008,Schruba2011,Bigiel2011}.
\citet{Bigiel2008} found a linear scaling relation between the SFR and
H$_2$ surface density.  Similarly, \citet{Gao2004} used emission from
HCN as a tracer of dense gas in galaxies and also found a linear
scaling relation with the IR luminosity and star formation rate.  In
nearby clouds ($<$1 kpc from the Sun), the relationship between gas
content and star formation can be accurately studied, where the gas
surface density is derived using dust extinction, and the star
formation rate is derived by counting young stellar objects (YSOs;
\citealt{Heiderman2010,Lada2010,Gutermuth2011}). These studies suggest
a steeper slope, $\sim 2$, of the SFR--gas relationship, compared with
extragalactic observations. But, above a certain surface density
threshold, the relationship becomes linear, suggesting that the
difference between the local and extragalactic SFR--gas relationships
is produced by the different fractions of dense gas traced by the
observations.

Most studies of external galaxies rely on a constant CO--to--H$_2$
conversion factor to convert the bright but optically thick $^{12}$CO
$J=1\to0$ line into a H$_2$ surface density.  In \citet{Pineda2013},
we used [C\,{\sc ii}], H\,{\sc i}, $^{12}$CO, and $^{13}$CO
observations to study the surface density distribution of the
different phases of the interstellar medium, including warm and cold
atomic gas, CO--dark H$_2$, and CO--traced H$_2$, across the Galactic
plane in a detail that currently is not possible for external
galaxies. We showed that the $X_{\rm CO}$ conversion factor varies
with Galactocentric distance as a consequence of the Galactic
metallicity gradient.  Additionally, we determined the distribution of
the CO--dark H$_2$ component (molecular gas that is not traced by CO
but by [C\,{\sc ii}];
\citealt{Madden1997,Madden2013,Langer2010,Langer2014a,Wolfire2010}),
which due to the lack of [C\,{\sc ii}] maps has only been studied in a
handful of nearby galaxies \citep[e.g.][]{Madden1997,Cormier2014}.  In
the present paper we analyze how the different phases of the ISM studied in
\citet{Pineda2013} are related to the star formation rate in the
Galactic plane, with the aim to reconcile the local and extragalactic
SFR--gas relationships.

This paper is organized as follows. In
Section~\ref{sec:star-formation-rate}, we use the observed free--free
emission as a function of Galactic longitude for $b=0\degr$ to
determine the distribution of the star formation rate in the Galaxy,
and in Section~\ref{sec:c-sc-ii-1} we determine the [C\,{\sc ii}]
luminosity distribution of the Milky Way based on the azimuthally
averaged [C\,{\sc ii}] emissivities derived in \citet{Pineda2013}
and the scale height of the [C\,{\sc ii}] emission derived by
\citet{Langer2014b}.  In Section~\ref{sec:comp-betw-sfr} we 
compare the SFR of the Galaxy with the [C\,{\sc ii}] emission arising
from the different ISM phases. In Section~\ref{sec:relat-betw-gas}, we
study the relationship between the SFR and the gas content of these
ISM phases, and in Section~\ref{sec:conclusions}, we summarize our
results.

\section{Star Formation Rate Distribution}
\label{sec:star-formation-rate}

The thermal radio continuum (free--free) emission arising from ionized
gas can be used to determine the number of Lyman continuum (Lyc)
photons, $N_{\rm Lyc}$, produced by ionizing stars.  $N_{\rm Lyc}$ can
then be used to determine the star formation rate required to maintain
a steady--state population of these stars. It has been shown that most
O stars in the Galaxy are located in (and ionize) extended low density
H\,{\sc ii} regions \citep{Mezger1978}, with only 10--20\% of O stars
being embedded in compact H\,{\sc ii} regions \citep{Mezger1976}. The
extended free-free emission observed in the Galactic plane originates
mostly from an ensemble of evolved H\,{\sc ii} regions whose
Str\"omgren spheres partially overlap. In the following, we refer to
this component of the ionized gas as the Extended Low Density Warm
Ionized Medium (ELDWIM).

We used the 1.4\,GHz free--free intensity distribution as a function
of Galactic longitude ($l$), and latitude ($b=0$), $I_{\rm
ff}(l,b=0)$, to derive the radial distribution of the star formation
rate in the Galactic plane. The free--free intensities are taken from
\citet{Guesten1982} and were derived by \citet{Mathewson1962} and
\citet{Westerhout1958} for the southern and northern parts of the
Galaxy, respectively. The frequency and angular resolution of these
data sets are (1.44\,GHz, 50\arcmin) for the \citet{Mathewson1962} and
(1.39\,GHz, 34\arcmin) for the \citet{Westerhout1958} data
sets\footnote{ The angular resolution of the \citet{Mathewson1962} and
\citet{Westerhout1958} data sets correspond to 40 pc and 50 pc at a
distance of 4\,kpc from the sun, respectively.}.  To isolate the
contribution from free--free emission in the observed radio continuum
distribution, \citet{Westerhout1958} and \citet{Mathewson1962}
subtracted the contribution from non--thermal synchrotron emission
using the 85.5\,MHz survey by \citet{Hill1958} (which is dominated by
non--thermal emission) assuming a flux density spectral index
($I\propto\nu^{\beta}$) for this emission mechanism of $-$2.6, between
85.5\,MHz and 1.4\,GHz.  The uncertainties in the assumed spectral
index of the non--thermal emission dominates the uncertainties in the
separation between thermal and non--thermal contributions to the 1.4
GHz emission. To compare the free--free distribution from
\citet{Westerhout1958} and \citet{Mathewson1962} with modern--day
data, we compared it with that derived for WMAP by \citet{Gold2011} at
23\,GHz using the Maximum Entropy Method with 1\degr\ angular
resolution. This data set includes the extended free--free emission
originating from the ELDWIM plus that from compact H\,{\sc ii}
regions.  We corrected for the difference in frequencies between the
two data sets and reduced the emission derived for WMAP by 50\% as
suggested by \citet{Alves2012}. We find that the longitudinal
distribution of the ELDWIM free--free emission derived for WMAP and
those derived by \citet{Westerhout1958} and \citet{Mathewson1962} are
in good agreement, with their intensities agreeing within 20--30\%. We
assumed uncertainties in the free--free distribution of 30\%. Note
that we prefer to use the free--free distribution from
\citet{Westerhout1958} and \citet{Mathewson1962} because point sources
have been subtracted (by identifying closed contours in their maps
having sizes close to that of their beam size), and therefore their
data set correspond to the free--free emission originating from the
ELDWIM. The contribution from compact H\,{\sc ii} in the derivation of
the SFR will be added in later in our analysis.  We first converted
the free--free intensity to the radial distribution of the azimuthally
averaged free--free emissivity. This emissivity can be converted into
the radial distribution of the star formation rate as described below.

We assumed that the galaxy is axisymmetric and can be divided in a set
of $N_{\rm rings}$ concentric rings of radius $R_i$ with each having an
azimuthally--averaged free--free emissivity $\epsilon_{\rm
ff}(R_i)$. The free--free intensity for a given line--of--sight
defined by its Galactic longitude $l_j$ is a linear sum over all rings
that it intersects,

\begin{equation}
 I_{\rm ff}(l_j)=\sum^{N_{\rm rings}}_{i=1} L_{ij}\epsilon_{\rm ff}(R_i),\,\, 1<j\leq M_{\rm LOS}, 
\end{equation}
where $L_{ij}$ describes the path length subtended over a
Galactocentric ring $R_i$ by a line-of-sight with Galactic longitude
$l_j$ and $M_{\rm LOS}$ is the number of sampled lines of sight. (For
an illustration of the adopted geometry see Fig. A1 in
\citealt{Pineda2013}.) This set of $M_{\rm LOS} \times N_{\rm rings}$
linear equations can in principle be inverted in the case where the
number of samples of the free--free intensity ($M_{\rm LOS}$) is equal
to the number of Galactocentric rings that produce the emission
($N_{\rm rings}$), resulting in the distribution of $\epsilon_{\rm
ff}(R)$ that reproduces the distribution of $I_{\rm ff}(l)$. In
practice, however, we do not expect to have an exact solution because
the Galaxy is not axisymmetric and the number of rings typically
available is not large enough to sample properly the $I_{\rm ff}(l)$
distribution.  We instead search for the distribution $\epsilon_{\rm
ff}(R)$ that minimizes the $\chi^2$ calculated between the observed
$I_{\rm ff}(l)$ distribution and that resulting from a modeled
$\epsilon_{\rm ff}(R)$ distribution.  Because the contribution from
outer galaxy rings to the free--free brightness temperature is likely
to be very small, there could be solutions that will have an
oscillatory behavior in the outer galaxy and still produce an
integrated intensity along the line of sight that is close to zero.
Additionally, rings close to the Galactic center are not well sampled
by the $I_{\rm ff}(l)$ distribution. Therefore, following
\citet{Assef2008,Assef2010}, we include a smoothing term $H$ in the
minimization that forces rings near the Galactic Center ($R_{\rm
gal}<$3\,kpc) and in the outer galaxy ($R_{\rm gal}>8.5$\,kpc) to be
close to zero.  We tested the effect of varying these limits in
$R_{\rm gal}$ in the solution of the $\chi^2$ minimization and found
that while the solution does not vary significantly for
$3$\,kpc$<R_{\rm gal}<8.5$\,kpc, using the smoothing term does
minimize the oscillatory behavior for the solution outside this range
in Galactocentric distance.   We therefore minimize the function

\begin{equation}
\label{eq:3}
G=\chi^2+\frac{H}{\eta^2},
\end{equation}
where $\eta$ is a free parameter that can be adjusted to keep the
values near the Galactic center and in the outer galaxy close to zero.
The $\chi^2$ function is given by
 
\begin{equation}
\chi^2=\sum_{j}^{M_{\rm LOS}} \left(\frac{I_{\rm ff}(l_j)-I_{\rm ff,\rm model}(l_j)}{\sigma(l_j)} \right )^2 ,
\end{equation}
where $\sigma(l_j)$ are the uncertainties in the free--free
intensity and $I_{\rm ff,\rm model}(l_j)$ is the free--free intensity predicted at $
l_j$ for a given radial emissivity distribution. The smoothing term is given by
\begin{equation}
H=\sum_{i_0<i<i_1} \epsilon_{\rm ff}(R_i)^2,
\end{equation}
where $R_{i_0}=3$\,kpc and $R_{i_1}=8.5$\,kpc define the range in
Galactocentric distance in which we expect to have most of the
contribution to the observed free--free intensity. 
Minimizing the function $G$ results in the set of equations

\begin{equation}
b_k=(a_{k,i}+c_{k,i})\epsilon_{\rm ff}(R_i),
\end{equation}
where
\begin{equation}
b_k=\sum^{M_{\rm LOS}}_{j} \frac{I_{\rm ff}(l_j) L_{k,j}}{\sigma(l_j)^2},
\end{equation}
\begin{equation}
a_{k,i}=\sum^{M_{\rm LOS}}_{j} \frac{L_{k,j} L_{i,j}}{\sigma(l_j)^2},
\end{equation}
and
\begin{equation}
c_{k,i}=\begin{cases}
1/\eta^2   &(i<i_0 \mbox{ or } i>i_1; \mbox{ and } k=i)\\
0          &(\mbox{otherwise}).\\
\end{cases}
\end{equation}

\begin{figure}[t]
  \centering
   \includegraphics[width=0.45\textwidth]{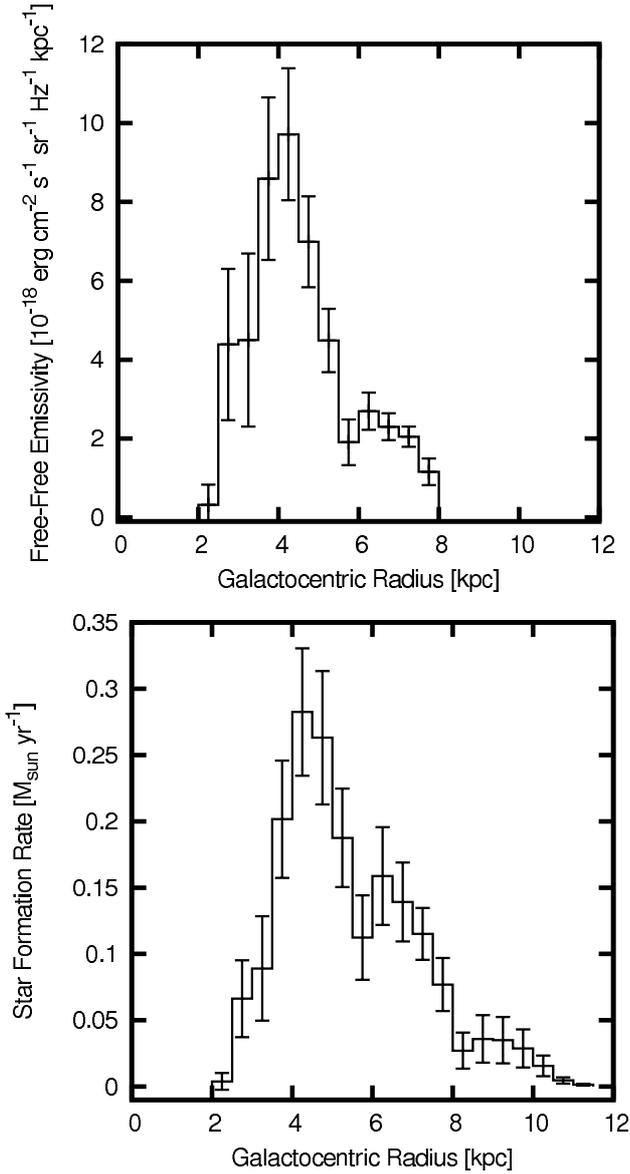}
      \caption{({\it upper panel}) Radial distribution of the
      free--free emissivity resulting from the average of those from
      the north and south portions of the Galaxy for $R_{\rm gal}>2$\,kpc. ({\it lower panel} )
      Radial distribution of the star formation rate derived following
      the method described in Section~\ref{sec:star-formation-rate}  for $R_{\rm gal}>2$\,kpc. }.
\label{fig:rc_distro}
\end{figure}

This set of $M_{\rm LOS} \times M_{\rm LOS}$ equations can be written
in vector--matrix form as
\begin{equation}
\vec{b}=(\vec{A+C})\vec{\epsilon_{\rm ff}},
\end{equation}
and the radial distribution of the free--free emissivity that best
 reproduced the observed $I_{\rm ff}(l)$ distribution can be
 determined by inverting the $\vec{A+C}$ matrix. The $I_{\rm ff}(l)$
 distribution used in our calculation is the average between that
 derived for the northern and southern parts of the Galaxy.  Note that
 because that all parameters involved in our calculation are linear,
 the solution we found by inverting the $\vec{A+C}$ matrix corresponds
 to the global minimum of the function $G$ (Equation~\ref{eq:3}).  In
 the upper panel of Figure~\ref{fig:rc_distro}, we show the resulting
 distribution of the free--free emissivity as a function of
 Galactocentric distance. The uncertainties in the emissivity
 distribution are given by the square root of the diagonal terms in
 the $(\vec{A+C})^{-1}$ matrix.  The resulting free--free emissivity
 distribution is in good agreement with that derived by
 \citet{Guesten1982}, which employed the same free--free data set but
 a different technique to determine the SFR.

%the free--free luminosity of each ring from its emissivity.
The free--free luminosity is related to the flux density observed at a
distance $D$ as,

\begin{equation}\label{eq:6}
L_{\rm ff}=4\pi D^2 S_{\rm ff},
\end{equation}
where $S_{\rm ff}$ is the flux density of the ring, which for
$D>>R_{\rm gal}$, is given by,

\begin{equation}
S_{\rm ff}=\frac{2\pi I_{\rm ff}(R_{\rm gal}) \Delta R_{\rm gal} R_{\rm gal}}{D^2}.
\end{equation}

Here, $I_{\rm ff}(R_{\rm gal})$ is the free--free continuum intensity
at the surface of a ring centered at a Galactocentric distance
$R_{\rm gal}$.  We assumed that the vertical distribution of the
free--free emissivity is Gaussian with a peak emissivity given by that
at $b=0$\degr\ and a FWHM $\Delta z_{\rm ff}$ of 260\,pc
\citep{Mezger1978}.  Note that pulsar dispersion observations
suggest the existence of a lower density ionized gas component that
extends even further vertically than the ELDWIM \citep[$\Delta z$ =
1\,kpc;][]{Kulkarni1987,Reynolds1989}.  $I_{\rm ff}(R_{\rm gal})$ is
related to the azimuthally averaged free--free emissivity at
$b=0$\degr, $\epsilon_{\rm ff}(R_{\rm gal})$, as

\begin{equation}
I_{\rm ff}(R_{\rm gal})=\frac{\sqrt{ 2\pi}}{2.354}\epsilon_{\rm ff}(R_{\rm gal})\Delta z_{\rm ff}.
\end{equation}

The free--free luminosity for a ring is therefore given by

\begin{equation}\label{eq:8}
L_{\rm ff}(R_{\rm gal})= \frac{2(2\pi)^{5/2}}{2.354} \epsilon_{\rm ff}(R_{\rm gal})\Delta z_{\rm ff} \Delta R_{\rm gal} R_{\rm gal}.
\end{equation}

We use the free--free luminosity distribution to calculate the number
of Lyc photons per second absorbed by the gas at Galactocentric
distance $R_{\rm gal}$.  The number of Lyc photons absorbed per second
by the ionized gas with electron temperature $T_{\rm e}$ and at
distance $D$, which emits free--free luminosity $L_{\rm ff}$ is given
by \citet{Mezger1978}

\begin{multline}
\label{eq:1}
\left[\frac{N_{\rm Lyc}(R_{\rm gal})}{\rm photons\,s^{-1}}\right]=\\
\frac{1.25}{(1+y^+)}6.32\times10^{52}\left[\frac{T_{\rm e}}{\rm 10^4 K} \right]^{-0.45}\left[\frac{\nu}{\rm GHz} \right]^{0.1}\left[\frac{L_{\rm ff}(R_{\rm gal})}{\rm 10^{27} erg\,s^{-1}\,Hz^{-1}} \right],
\end{multline}
where $y^+=0.08$ corrects for the abundance of singly ionized He and
the factor 1.25 accounts for the number of Lyc photons directly
absorbed by dust (we assumed that the number of Lyc photons that can
escape the Galactic disk is negligible). We assumed an electron
temperature, $T_{\rm e}=7000$\,K \citep{Mezger1978}.  We also include
in the radial distribution of the number of Lyc photons the
contribution from Lyc photons ionizing compact H\,{\sc ii} regions
\citep{Smith1978}. This contribution is small compared with that from
the ELDWIM \citep{Mezger1978}, but it extends over a wider range of
Galactocentric distances.

The star formation rate required to maintain a steady--state
population of ionizing stars that produces the observed number of
Lyman continuum photons per second can be determined using the
relationship SFR[M$_{\odot}\,{\rm yr}^{-1}]=7.5\times10^{-54} N_{\rm
Lyc} [{\rm photons}\,{\rm s}^{-1}]$ estimated by \citet{Chomiuk2011}
assuming a Kroupa initial mass function (\citealt{Kroupa2003}), Solar
metallicity, and considering population models of O stars. The
resulting radial distribution of the star--formation rate in the
Galactic plane for $R_{\rm gal}>2$\,kpc is shown in the lower panel of
Figure~\ref{fig:rc_distro}. Integrating the star formation rate across
the galaxy, we obtain a total star formation rate for the Milky Way of
$1.8\pm0.1$ M$_{\odot}$\,yr$^{-1}$, in excellent agreement with the
$1.9\pm0.4$ M$_{\odot}$\,yr$^{-1}$ estimated by \citet{Chomiuk2011}
using several independent methods.  Note that our estimation of the
total SFR of the Milky Way does not include the contribution from the
Galactic center. The uncertainties in our estimation the Galactic SFR
are likely a lower limit, as these are calculated by propagating the
errors of the radial SFR distribution shown in
Figure~\ref{fig:rc_distro}, and do not account for uncertainties
resulting from the conversion from $N_{\rm Lyc}$ to SFR and the
assumption that the Galaxy is axisymmetric. \citet{Guesten1982}
studied the effect on the radial distribution of $N_{\rm Lyc}$ of
considering the spiral arm structure of the Milky Way, which is a
barred spiral (SBc) galaxy. They show that within the accuracy of
their estimation of $N_{\rm Lyc} (R_{\rm gal})$, accounting for spiral
arms does not have a significant effect on the derived contribution
compared to that derived under the assumption of azimuthal symmetry.

The free--free emission originates from H\,{\sc ii} regions that
are produced by massive stars ($M>15$\,M$_{\odot}$), and therefore
traces the star formation averaged over 3--10 Myr, which is the
lifetime of O stars \citep[e.g.][]{Kennicutt2012}. On the other hand,
the FUV that heats the neutral ISM, which is later cooled by [C\,{\sc
ii}], also originates from B--stars with masses above 5\,M$_{\odot}$
\citep{Parravano2003}, thus tracing the SFR over a longer time scale
(100\,Myr).  Note that the conversion between $N_{\rm Lyc}$ and the
star formation rate needs to be evaluated over spatial scales that are
sufficiently large that the IMF is well sampled \citep[see
e.g.][]{Vutisalchavakul2013}. By dividing the Milky Way in a set of
rings, we are averaging free--free continuum emission, and therefore
evaluating the star formation rate, over scales between 10 to 30
kpc$^2$, which correspond to the range of the ring areas considered
here. For such areas, the IMF is expected to be fully sampled
\citep{Kennicutt2012}.  Another assumption that we need to make to use
the Lyc continuum photons to trace the star formation rate is a
continuous star formation over the scales at which this SFR
calibration works (3--10\,Myr). This assumption is justified as the
SFR in the Milky Way has been suggested to be nearly continuous over
the last $\sim$8\,Gyrs \citep{Noh1990,Rocha-Pinto2000,Fuchs2009} with
some variation on a period of $\sim$400\,Myr \citep{Rocha-Pinto2000},
thus the timescales for variation in the SFR are much longer than that
on which the Lyc photons trace the SFR.

Note that the conversion from the number of Lyc photons to the star
formation rate is sensitive to metallicity, which is known to vary
with Galactocentric distance \citep[e.g][]{Rolleston2000}. These
effects have been studied with evolutionary synthesis models by
\citet{Schraerer1998} and \citet{Smith2002} \citep[see
also][]{Kennicutt2012}.  According to the models of \citet{Smith2002},
for continuous star formation at a rate of 1\,M$_{\odot}$\,yr$^{-1}$,
the resulting ionizing photon luminosity varies by only 30\% for a factor of 5
change in metallicity.  In the range where most of the free--free
emissivity is observed in the Galactic plane, 3\,kpc$<R_{\rm
gal}<$10\,kpc, the metallicity varies by a factor of 1.5 from its
average value (1.5\,Z$_{\odot}$), for the metallicity gradient fitted
by \citet{Rolleston2000}. We therefore expect that the dependence of the
conversion from $N_{\rm Lyc}$ and SFR with metallicity is negligible
in the Galactic plane, and we use the conversion for solar metallicity
without any further correction.

\section{[C\,{\sc ii}] Luminosity of the Milky Way}
\label{sec:c-sc-ii-1}

\begin{figure}[t]
  \centering
   \includegraphics[width=0.48\textwidth]{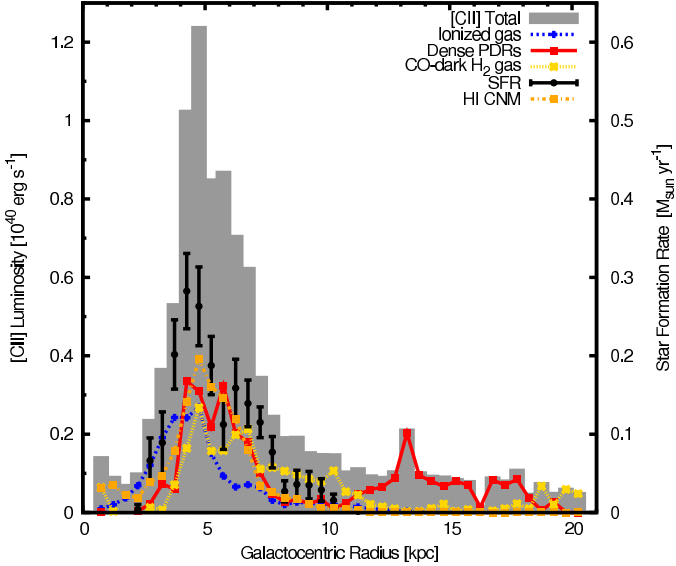}
      \caption{The [C\,{\sc ii}] luminosity of the Milky Way as a
      function of Galactocentric radius. The estimated contributions
      to the [C\,{\sc ii}] luminosity from gas associated with dense
      PDRs, ionized gas, CNM H\,{\sc i} gas, and CO--dark H$_2$ gas
      are also shown. Typical uncertainties are 5\% for the total
      [C\,{\sc ii}] luminosity and the contribution from H\,{\sc i}
      gas, 10\% for the contribution from CO--dark H$_2$ gas, and we
      assume 30\% for the contribution from ionized gas. The black
      dots (and error bars) represent the radial distribution of the
      star--formation rate derived in
      Section~\ref{sec:star-formation-rate}.}
\label{fig:cii_sfr_distro} 
\end{figure}

We estimated the radial luminosity distribution of the different ISM
components contributing to the observed [C\,{\sc ii}] emissivity in
the Galactic plane (dense PDRs, cold H\,{\sc i}, CO--dark H$_2$, and
ionized gas).  We determined the [C\,{\sc ii}] luminosity distributions
using the emissivity distributions derived in \citet{Pineda2013}
following the same procedure used to determine the free--free
luminosity,
\begin{equation}
L_{\rm [CII]}(R_{\rm gal})= \frac{2(2\pi)^{5/2}}{2.354} \epsilon_{\rm [CII]}(R_{\rm gal})\Delta z_{\rm [CII]} \Delta R_{\rm gal} R_{\rm gal}.
\end{equation}
Following the results presented by \citet{Langer2014b}, we assume a
scale height of the Galactic disk with a FWHM of 172\,pc for the
CO--dark H$_2$ gas, 212\,pc for the cold H\,{\sc i} gas
\citep{Dickey1990}, and 130\,pc for the dense PDR component.   We
considered the diffuse ionized component discussed by
\citet{Kulkarni1987} and \citet{Reynolds1989}, that has a FWHM of
1\,kpc. This scale height is consistent with that constrained by
pulsar dispersion in the NE2001 model \citep{Cordes2003}.  The
resulting [C\,{\sc ii}] luminosity distributions are shown in
Figure~\ref{fig:cii_sfr_distro}. We also show the radial distribution
of the total [C\,{\sc ii}] luminosity of the Milky Way, which is the
sum of the luminosity from all contributing ISM components.  The total
[C\,{\sc ii}] luminosity of the Milky Way and the contributions to
that from different ISM components are listed in
Table~\ref{tab:fits_sfr_cii}.   The [C\,{\sc ii}] luminosity of
the different ISM components is the sum of that for each
Galactocentric ring, 
\begin{equation}
L_{\rm [CII]}=\sum^{N}_{i} L_{\rm [CII]}(R^i_{\rm gal}).
\end{equation}
The ISM components that contribute to the [C\,{\sc ii}] luminosity of
the Milky Way have roughly comparable contributions: dense PDRs
(30\%), cold H\,{\sc i} (25\%), CO--dark H$_2$ (25\%), and ionized gas
(20\%).

 Note that the contribution from ionized gas to the total [C\,{\sc
ii}] luminosity of the Galaxy is much larger than its $\sim$4\%
contribution to the total [C\,{\sc ii}] emissivity at $b=0$\degr\,
estimated by \citet{Pineda2013}, using the NE2001 \citep{Cordes2002}
model of the distribution of ionized gas in the Galaxy.  The [C\,{\sc
ii}] emission associated with ionized gas is expected to be diffuse
but much more vertically extended compared with the other [C\,{\sc
ii}]--emitting ISM components. Therefore, since the scale height of
the ionized gas is much larger than that of either atomic and
molecular gas, the [C\,{\sc ii}] luminosity associated with ionized
gas becomes significant despite its emissivity at $b=0$\degr\ being
relatively low. The larger contribution of ionized gas to the total
[C\,{\sc ii}] luminosity of the Milky Way is in agreement with the
suggestion by \citet{Heiles1994}, that ionized gas could be a
significant contributor to the [C\,{\sc ii}] intensity observed by
COBE.  Note that the NE2001 model assumes a distribution of electron
densities that is smooth on the order of kpc scales.  Therefore, our
estimate of the contribution of the ionized gas to the total [C\,{\sc
ii}] luminosity should be considered as a lower limit, in the case
that the ELDWIM is clumpy.

   \begin{figure}[t]
  \centering
   \includegraphics[width=0.48\textwidth]{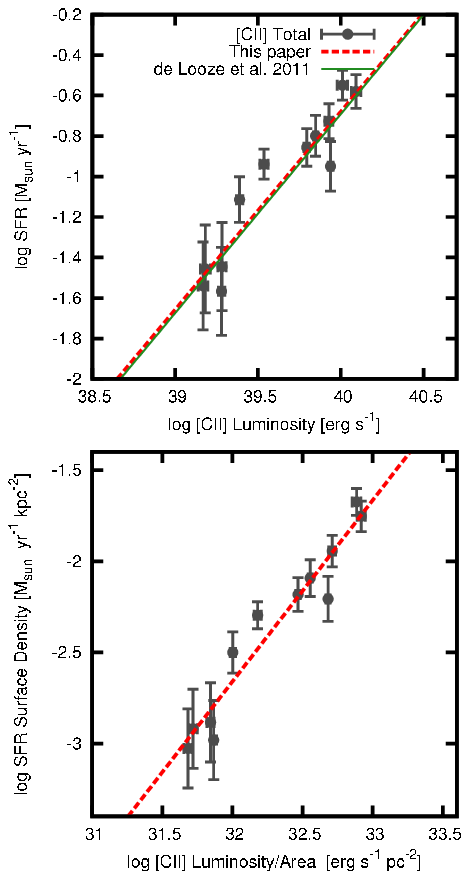}
      \caption{({\it upper panel}) Comparison between the star
      formation rate and the observed [C\,{\sc ii}] luminosity for
      different rings in the Milky Way.  The dashed straight line
      represents a fit to the data while the solid straight line is
      the relationship between the star formation rate and [C\,{\sc
      ii}] luminosity fitted by \citet{deLooze2011} for a set of
      nearby galaxies. ({\it lower panel}) Comparison between the star
      formation rate surface density and the [C\,{\sc ii}] luminosity
      per unit area. The dashed straight line represents a fit to the
      data.  }
\label{fig:sfr_vs_cii}
\end{figure}

 We estimate a total [C\,{\sc ii}] luminosity of the Milky Way of
$L^{\rm MW}_{\rm [CII]}=1.0\times10^{41}$ erg s$^{-1}$. This value of
$L^{\rm MW}_{\rm [CII]}$ is about half that estimated by
\citet{wright1991} of $L^{\rm MW}_{\rm [CII]}=1.95\times10^{41}$ erg
s$^{-1}$ using COBE data.  \citet{wright1991} derived the [C\,{\sc
ii}] luminosity using $L^{\rm MW}_{\rm
[CII]}$=4$\pi$R$_\odot^2[(1.35$\,sr$)\langle I_{\rm [CII]}\rangle$],
where $\langle I_{\rm [CII]} \rangle =1.7\pm0.07\times
10^{-5}$\,erg\,cm$^{-2}$\,s$^{-1}$\,sr$^{-1}$ is the mean [C\,{\sc
ii}] intensity observed by COBE. The average [C\,{\sc ii}] intensity
is derived over a solid angle of 1.35\,sr, which is the result of the
integral of a source function that is derived by fitting the
distribution of dust continuum emission observed by COBE. Because the
spatial distribution of the dust continuum distribution is not
necessarily the same as that of [C\,{\sc ii}], this might be the
origin of the discrepancy.  To understand the source of the difference
with COBE we analyzed the GOT C+ data using an approach similar to
that used for the COBE data as follows.  We constructed a synthetic
map of the Milky Way in [C\,{\sc ii}] using the radial distribution of
the [C\,{\sc ii}] emissivity shown in Figure~4 in \citet{Pineda2013}
and assuming a FWHM of the [C\,{\sc ii}] emission of 172\,pc. This map
is then convolved with a 7\degr\ beam to simulate the emission
observed by COBE.  Since \citet{wright1991} do not provide the
functional form of the source function, we used the publicly available
FIRAS data (which was reprocessed by \citealt{Bennett1994}) to
estimate the mean [C\,{\sc ii}] intensity observed by COBE.
Integrating the [C\,{\sc ii}] emission in the COBE map using a
5$\sigma$ threshold, where $\sigma=7\times
10^{-7}$\,erg\,cm$^{-2}$\,s$^{-1}$\,sr$^{-1}$, yields an average
[C\,{\sc ii}] intensity $\langle I_{\rm [CII]}
\rangle=1.8\times10^{-5}$\,erg\,cm$^{-2}$\,s$^{-1}$\,sr$^{-1}$, which
is consistent with that published by \citet{wright1991}.  In our
simulated map, the $5\sigma$ limit yields an averaged intensity of
$\langle I_{\rm [CII]}
\rangle=1.6\times10^{-5}$\,erg\,cm$^{-2}$\,s$^{-1}$\,sr$^{-1}$, which
corresponds, using the equation used by \citet{wright1991}, to $L^{\rm
MW}_{\rm [CII]}=1.8\times 10^{41}$\,erg\,s$^{-1}$, in excellent
agreement with the [C\,{\sc ii}] luminosity suggested by COBE.
Therefore the difference between the [C\,{\sc ii}] luminosities
derived from GOT\,C+ and COBE likely arises from the different methods
and assumptions used regarding the volume distribution of [C\,{\sc
ii}] emission.  Note also that there may be some uncertainty on the
absolute flux calibration of the COBE/FIRAS instrument, as suggested
by the discrepancy between absolute [C\,{\sc ii}] fluxes observed by
COBE/FIRAS and FILM/IRTS \citep{Makiuti2002}, with the latter
instrument showing $\sim50$\% lower flux.  We suggest that our
[C\,{\sc ii}] luminosity estimation is better suited for comparison
with external galaxies than that presented by \citet{wright1991}.

\section{[C\,{\sc ii}] as an indicator of the star formation rate}
\label{sec:comp-betw-sfr}

   \begin{figure*}[t]
  \centering
   \includegraphics[width=0.75\textwidth]{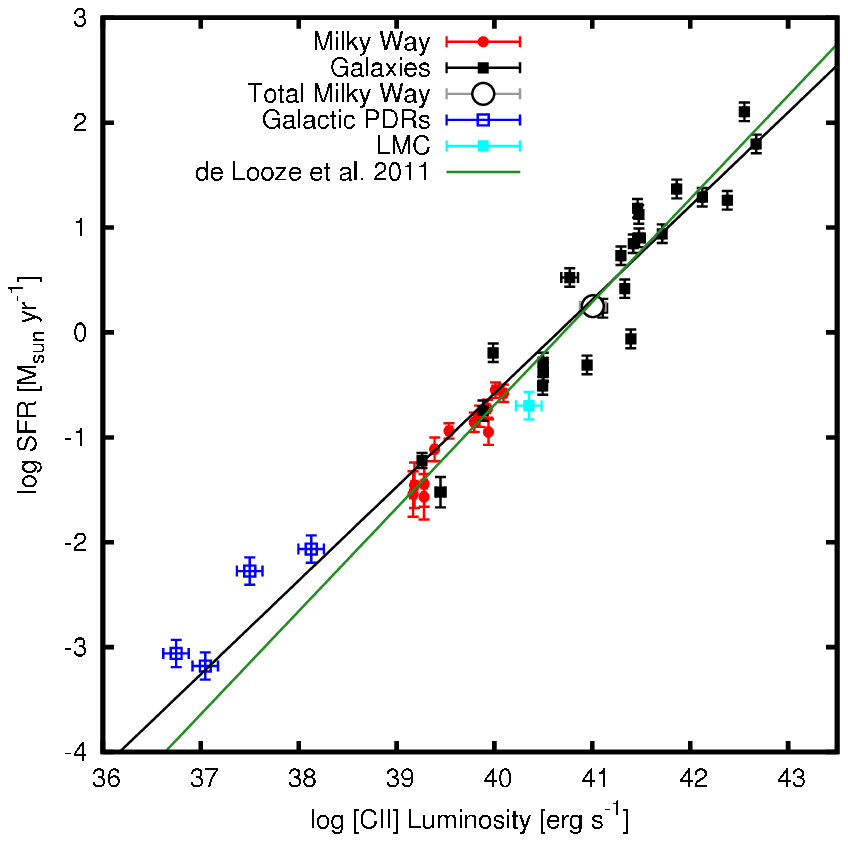}
   \caption{Star formation rate as a function of [C\,{\sc ii}]
   luminosity for different rings in the Galaxy, the LMC, the galaxies
   studied by \citet{deLooze2011}, and for individual Galactic photon
   dominated regions. The [C\,{\sc ii}] luminosity of the LMC is taken
   from \citet{Rubin2009} and the SFR from \citet{Harris2009}. The
   Galactic PDR data points include Orion ($L_{\rm [CII]} $ from
   \citealt{Stacey1993} and SFR from \citealt{Lada2010}), M17 ($L_{\rm
   [CII]} $ from \citealt{Stacey1991} and SFR from
   \citealt{Lada2010}), Rosette ($L_{\rm [CII]} $ from
   \citealt{Schneider1998} and SFR from \citealt{Chomiuk2011}), and
   Carina ($L_{\rm [CII]} $ from \citealt{Brooks2003} and SFR from
   \citealt{Chomiuk2011}).  The [C\,{\sc ii}] luminosities and SFRs
   for this sample are strongly correlated over 6 orders of
   magnitude. }
\label{fig:galaxies_sfr}
\end{figure*}

   In Figure~\ref{fig:cii_sfr_distro}, we show the radial distribution
   of the SFR together with that of the total [C\,{\sc ii}] luminosity
   and the different contributing ISM  components.  The SFR and total
   [C\,{\sc ii}] luminosity distributions are very similar with most
   of the star--formation taking place and the [C\,{\sc ii}] emission
   being produced between 4 and 10\,kpc. Both distributions show a
   peak at $\sim$4.5\,kpc from the Galactic center.

In the top panel of Figure~\ref{fig:sfr_vs_cii}, we show the SFR as a
   function of the [C\,{\sc ii}] luminosity between 4 and 10\,kpc for
   bins in Galactocentric radius of 0.5\,kpc width.  As discussed
   in \citet{Pineda2013}, we find that the bin width of 0.5\,kpc over
   the range of Galactocentric distance used here allows us to study
   radial variations of the [C\,{\sc ii}] emission while ensuring that
   each ring is sufficiently sampled by the GOT\,C+ data.  The
   uncertainties in the [C\,{\sc ii}] luminosity are derived by the
   propagation of the errors in the integrated intensities measured
   for all samples of a given ring \citep[see][]{Pineda2013}. We
   fitted a straight line to the data using the orthogonal bi-variate
   error and intrinsic scatter method (BES; \citealt{Akritas1996}).
   The result of the fit
\begin{equation}
\log {\rm SFR} = m \log L_{\rm [CII]}+b, 
\end{equation}
where $m$ is the slope and $b$ the intercept, is shown in the top
   panel of Figure~\ref{fig:sfr_vs_cii} and listed in
   Table~\ref{tab:fits_sfr_cii}. We also include in
   Table~\ref{tab:fits_sfr_cii} the standard deviation, which measures
   the dispersion of the observed SFRs from the fit, and the
   Spearman's rank coefficient, $\rho$, which evaluates the strength
   of the correlation\footnote{ A strong correlation or
   anti--correlation is characterized by values of $\rho$ approaching
   $+1$ or $-1$, respectively, while a lack of correlation is
   characterized by values $\rho$ approaching 0.  }.  We also show
   the fit to the relationship between SFR and [C\,{\sc ii}]
   luminosity for a sample of nearby galaxies presented by
   \citet{deLooze2011}, who sampled 24 nearby star--forming galaxies,
   including starbursts and low-ionization nuclear emission-line
   region (LINERs) galaxies.  The SFR and the total [C\,{\sc ii}]
   luminosity we determined from the GOT\,C+ data are well correlated
   ($\rho=0.91$), with a slope that is in very good agreement with
   that estimated by \citet{deLooze2011}. The dispersion of the
   observed SFRs from the fit is 0.14\,dex which is smaller than the
   0.27\,dex found by \citet{deLooze2011} for their sample of
   galaxies, which could be an indication that the variation in the
   ISM conditions (e.g. densities, temperatures, FUV fields) within
   the Milky Way is smaller than that among the galaxies in their
   sample. A factor that could enhance the observed correlation is
   the fact that both the [C\,{\sc ii}] luminosity and SFR are
   functions of the area of a given ring, which is in turn a function
   of Galactocentric distance. In the bottom panel of
   Figure~\ref{fig:sfr_vs_cii} we show the SFR surface density
   (SFR/area) as a function of the [C\,{\sc ii}] luminosity divided by
   area. We see that the correlation still holds, with a Spearman's
   rank coefficient of 0.92, thus indicating that the ring area is not
   the source of the correlation seen above. The straight line
   fit, $\log$($\Sigma_{\rm
   SFR}$\,[M$_\odot$\,yr$^{-1}$\,kpc$^{-2}$)]=(0.99$\pm$0.06)$\log(L_{\rm
   [CII]}/{\rm Area}$\,[erg s$^{-1}$\,pc$^{-2}$]$-$(34.50$\pm$1.80),
   also shows that the slope is very similar to that of derived for
   the correlation between [C\,{\sc ii}] and SFR.

\begin{table*}                                            
\caption{Correlation between the [C\,{\sc ii}] luminosity of the different
contributing ISM components and the SFR for different rings in the Galactic plane obtained using BES fits}
\label{tab:fits_sfr_cii}					    
\centering						    
\begin{tabular}{l c c c c c }				    
\hline\hline
Variable  &  \multicolumn{2}{c}{$\log {\rm  SFR} = m \log L_{\rm [CII]}+b$} & Dispersion& Spearman's rank coeff.  &Total Luminosity    \\
 & $m$  & $b$  & dex &  &$10^{40}$ erg s$^{-1}$       \\
\hline
 
[C\,{\sc ii}] Total           & 0.98$\pm$0.07 &-39.80$\pm$2.94 & 0.14 & 0.91 & 10.1  \\
			     				           
 [C\,{\sc ii}] (PDRs)        & 0.74$\pm$0.07 &-30.00$\pm$2.81  & 0.12 & 0.88 & 3.2  \\
 			     				           
[C\,{\sc ii}] (Electrons)    & 0.91$\pm$0.06 &-36.30$\pm$2.36  & 0.11 & 0.94 & 1.9  \\
 			     				           
[C\,{\sc ii}] (HI\,CNM)      & 0.65$\pm$0.07 &-26.50$\pm$2.75  & 0.14 & 0.86  & 2.5  \\
 			     				           
[C\,{\sc ii}] (Dark\,Gas)     & 2.12$\pm$0.37 &-84.10$\pm$14.30 & 0.21 & 0.84  & 2.5  \\
\hline
\end{tabular}						    
\end{table*} 

   In Figure \ref{fig:galaxies_sfr}, we show the SFR as a function of
   [C\,{\sc ii}] luminosity for rings in the Galaxy, the LMC, the
   Galaxies studied by \citet{deLooze2011}, and Galactic photon
   dominated regions.  To determine the SFR in their sample of
   Galaxies, \citet{deLooze2011} used the FUV+24\,$\mu$m emission,
   while the SFR of Galactic star--forming regions was determined
   using YSO counts \citep{Chomiuk2011,Lada2010}. The different
   methods used for determining the SFR are likely one of the main
   reasons for the large scatter (0.3\,dex) in the relationship
   \citep{Vutisalchavakul2013}. Nevertheless, we see that the SFR and
   [C\,{\sc ii}] luminosity are correlated ($\rho=0.98$) over 6 orders
   of magnitude. We fitted a straight line to the combined data
   resulting in the relationship,
   $\log$(SFR[M$_\odot$\,yr$^{-1}$])=(0.89$\pm$0.04)$\log(L_{\rm
   [CII]}$[erg s$^{-1}])-(36.3\pm1.5)$.  The good correlation between
   the [C\,{\sc ii}] luminosity and SFR suggests that [C\,{\sc ii}] is
   a good tracer of the star formation activity in the Milky Way, as
   well as in nearby galaxies  dominated by star formation.

\begin{figure}[t]
  \centering
   \includegraphics[width=0.5\textwidth]{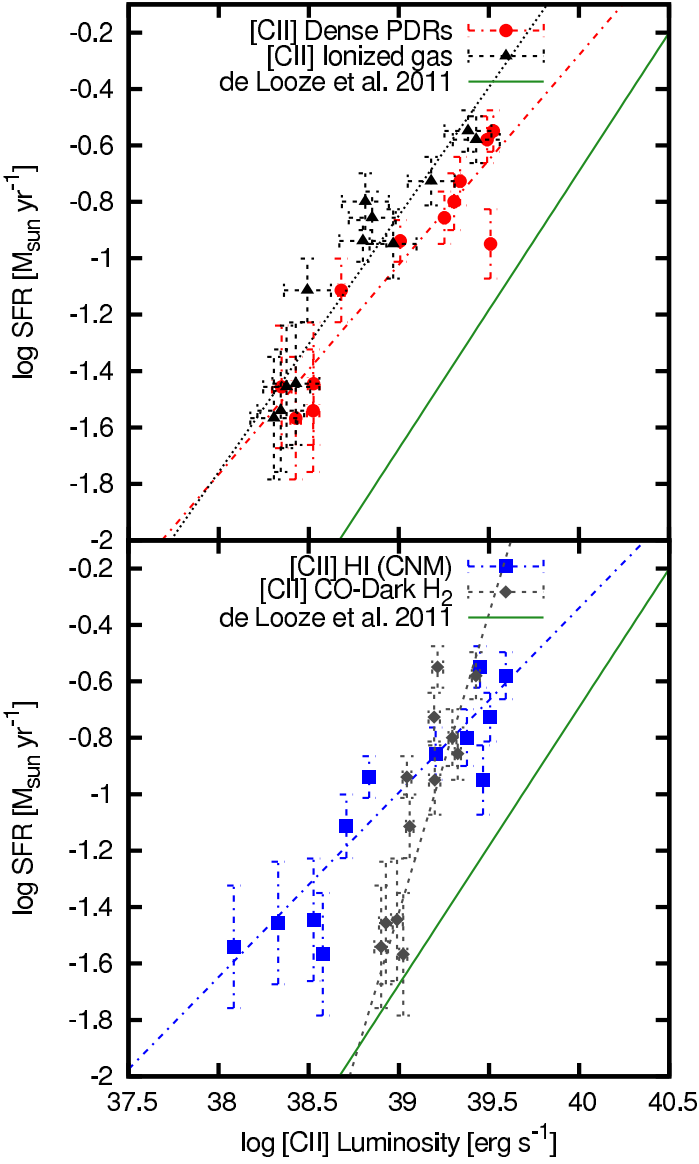}
      \caption{ ({\it upper panel}) Comparison between the star
      formation rate and the contribution to the [C\,{\sc ii}]
      luminosity from PDRs and electrons, and ({\it bottom panel})
      from cold H\,{\sc i} and CO--dark H$_2$ gas. In both panels the
      dashed straight lines represent fits to the data and the solid
      straight line is the relationship between the star formation
      rate and [C\,{\sc ii}] luminosity fitted by \citet{deLooze2011}
      in a set of nearby galaxies.  Typical uncertainties are 5\% for
      the total [C\,{\sc ii}] luminosity and the contribution from
      H\,{\sc i} gas, 10\% for the contribution from CO--dark H$_2$
      gas, and  30\% for the contribution from ionized gas.
      }
\label{fig:cii_sfr_phases}
\end{figure}

We also investigated the relationship between the SFR and [C\,{\sc
ii}] luminosity emerging from the different contributing ISM
components. In Figure~\ref{fig:cii_sfr_phases} we compare the SFR
versus the [C\,{\sc ii}] luminosity arising from dense PDRs, cold
H\,{\sc i} gas, ionized gas, and CO--dark H$_2$ gas for bins in
Galactocentric radius of 0.5\,kpc. We list the results of straight
line fits in Table~\ref{tab:fits_sfr_cii}, and we show the resulting
fits in Figure~\ref{fig:cii_sfr_phases}. For comparison we also
include the fit for nearby galaxies by \citet{deLooze2011}.  The
[C\,{\sc ii}] luminosity for the different contributing components are
well correlated with the SFR ($\rho=0.86-0.94$) with dispersions in
the 0.1--0.2\,dex range.  The slope of the relationships
between the star formation rate and the [C\,{\sc ii}] luminosity
arising from PDRs and ionized gas have slopes that are similar to that
found for the relationship involving the total [C\,{\sc ii}]
luminosity. For the relationship involving the cold H\,{\sc i} gas,
the relationship is somewhat shallower, while for the relationship
involving the CO-dark H$_2$ gas, the slope is significantly
steeper. At larger Galactocentric distances ($8-10$\,kpc), where there
is little star formation, the [C\,{\sc ii}] luminosity associated with
H\,{\sc i} is a weak contributor, while CO--dark H$_2$ dominates the
[C\,{\sc ii}] luminosity. The [C\,{\sc ii}] luminosity arising from
PDRs, ionized gas, and atomic gas show an steeper increase at smaller
Galactocentric distances compared to that originating from the
CO--dark H$_2$ gas. This increase in the [C\,{\sc ii}] luminosity is a
result of the combined effect of an increased column density and
thermal pressure. The column density of the CO--dark H$_2$, however,
stays constant between 4 and 10\,kpc \citep{Pineda2013}, as the
increased H$_2$ column densities, which is associated with a greater
dust shielding, enhances the abundance of CO. The constant CO--dark
H$_2$ column density results in a shallower increase in the [C\,{\sc
ii}] luminosity for reduced Galactocentric distances, which in turn
results in an steeper relationship with the star formation rate
compared with that involving the other ISM contributing components.  
The difference between the slopes of the different ISM contributing
components to the [C\,{\sc ii}] luminosity suggests that the observed
correlation between the total $L_{\rm [CII]}$ and SFR is the result of
the combined emission of the different contributing ISM phases.  Most
FUV photons produced by young stars are involved in heating the
gas. As the gas becomes warmer, the $^2P_{3/2}$ level of C$^+$ can be
more readily populated by collisions, and the rate of emission of
[C\,{\sc ii}] photons is increased.  Therefore, the [C\,{\sc ii}]
emission is indirectly related to the process of star formation,
regardless from which ISM phase it emerges.

In thermal equilibrium, the gas heating, which is dominated by the
photo--electric effect of FUV photons in dust grains and is thus
proportional to the star formation rate, is expected to be balanced by
the gas cooling, which is dominated by [C\,{\sc ii}] emission  (except
in the brightest and densest PDRs, where [O\,{\sc i}] emission also
contributes, and in ionized gas regions where excitation of low lying
electronic states of trace species such as O$^{++}$ and N$^{+}$
dominates).  For Gaussian distributions of the [C\,{\sc ii}] intensity
and H column density, we can calculate the midplane cooling rate per
hydrogen nucleus for a given Galactocentric radius as $\Lambda(R_{\rm
gal})= L_{\rm [CII]}(R_{\rm gal})/M_{\rm H}(R_{\rm gal})$, where
$L_{\rm [CII]}(R_{\rm gal})$ is the [C\,{\sc ii}] luminosity and
$M_{\rm H}(R_{\rm gal})$ is the H mass of a given ring.  To calculate
the cooling rate, we used the luminosity distribution of the different
ISM components presented in Figure~\ref{fig:cii_sfr_distro}, and we
estimated their H mass from the surface density distributions
presented in \citet{Pineda2013}, with the exception of the dense PDR
component (see below). In the upper panel of Figure~\ref{fig:cooling}
we compare the cooling rate from the different ISM components with the
star formation rate for different Galactocentric radii. As expected,
we see a good correlation between $\Lambda$ and the star formation
rate, with the different ISM components having relationships with the
SFR that have similar slopes.

We can understand the slope of the relationship between the cooling
rate of the different ISM components considering the distribution of
$\Lambda$ as a function of Galactocentric radius. In the lower panel
of Figure~\ref{fig:cooling} we see that the cooling rate increases
with decreasing Galactocentric radius. It can be shown that for
[C\,{\sc ii}] in the optically thin and subthermal regime, the cooling
rate per H nuclei is proportional to the C$^+$ to H abundance ratio,
the collisional rate coefficient, the H volume density, and is a
function of the kinetic temperature
\citep{Wiesenfeld2014,Goldsmith2012}.  Thus, the increase in the
cooling rate for decreasing Galactocentric radius is produced by
volume density and abundance gradients. This trend is exemplified by
the straight lines seen for the CO--dark H$_2$ and cold H\,{\sc i}
(CNM) components which are a reflection of the adopted thermal
pressure gradient from \citet{Wolfire2003}, which is used to calculate
the volume density assuming a constant kinetic temperature, and the
carbon abundance gradient from \citet{Rolleston2000}. Both the volume
density and the C$^+$ relative abundance increase by a factor $\sim$3
going from 10\,kpc to 4\,kpc.  Therefore, our results suggest that the
slope of the relationship between the cooling rate and the star
formation rate is a result of a combined effect of the volume density
and carbon abundance gradients in the Galactic plane.

\begin{figure}[t]
  \centering
   \includegraphics[width=0.48\textwidth]{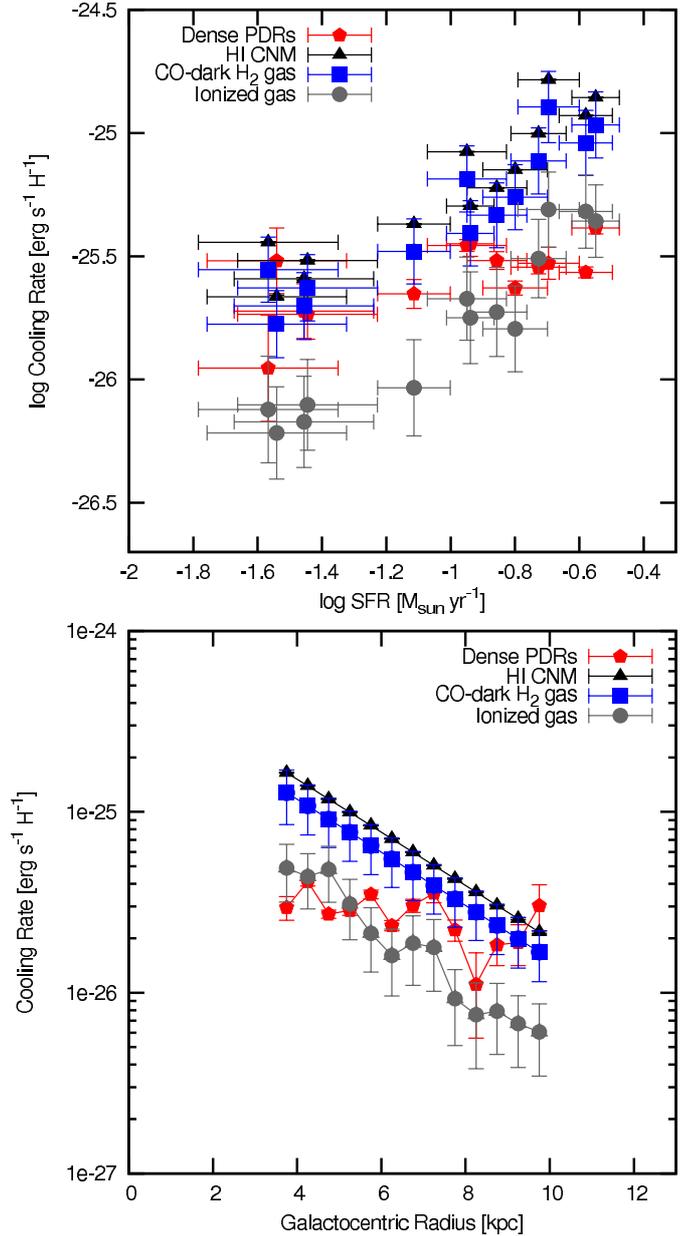}
   \caption{({\it upper panel}) The [C\,{\sc ii}] cooling rate per H
   atom as a function of the star formation rate for line--emitting
   gas associated with dense PDRs, cold H\,{\sc i} gas, CO--dark H$_2$
   gas, and ionized gas. ({\it lower panel}) The [C\,{\sc ii}] cooling
   rate as a function of Galactocentric radius for the different
   contributing ISM phases discussed in the text.  }
\label{fig:cooling}
\end{figure}

Note that we estimated the H mass distribution associated with dense
PDRs using $^{12}$CO and $^{13}$CO observations using the method
described in \citet{Pineda2013} for pixels in the position--velocity
map where [C\,{\sc ii}], $^{12}$CO, and $^{13}$CO are detected. The
derived column density does not necessarily represent the column of
gas associated with [C\,{\sc ii}] emission as there might be gas that
is shielded from the FUV photons where most of the carbon is in the
form of CO. Therefore, we consider the derived cooling rate for dense
PDRs as a lower limit. This effect should also vary as a function of
Galactocentric radius, as the average gas column density (and the
fraction of shielded gas) increases with Galactocentric distance (see
Figure~\ref{fig:surf_dense}). This effect might explain the flatter
slope seen in the cooling rate as a function of Galactocentric
distance.

\section{The Relationship between Gas content and Star--formation in the disk of the Milky Way}
\label{sec:relat-betw-gas}

Starting with the work by \citet{Schmidt1959} and \citet{Kennicutt98},
numerous studies have focused on the relationship between the star
formation activity and gas content in galaxies. The relationship
between star formation rate and gas surface density has been evaluated
over a wide range of spatial scales including entire galaxies, within
resolved disks of nearby galaxies
\citep[e.g.][]{Bigiel2008,Bigiel2011,Schruba2011,Ford2013}, and nearby molecular clouds
\citep[e.g.][]{Heiderman2010,Lada2010,Gutermuth2011}. The star
formation rate and total hydrogen (H+2H$_2$) surface densities have
been found to be correlated with a slope of 1.4 \citep{Kennicutt98},
while a linear relationship has been found between the SFR and
molecular gas surface density in resolved disks of galaxies
\citep{Bigiel2008}. A linear relationship is also found when the
emission from dense gas tracers (e.g. HCN) in galaxies is compared
with the star formation rate \citep{Gao2004}.  Most of the studies of
external galaxies rely on a CO--to--H$_2$ conversion factor ($X_{\rm
CO}$) to convert the optically thick $^{12}$CO $J=1\to0$ line into
H$_2$ surface densities, as observing times required for mapping the
nearly optically thin but weaker $^{13}$CO $J=1\to0$ line are
prohibitive.  Additionally, there are not many [C\,{\sc ii}]
observations in galaxies where the CO--dark H$_2$ gas contribution can
be studied.  In nearby clouds it is possible to estimate accurately
the SFR by counting YSOs and the gas surface density by using dust
extinction mapping. Thus the determination of the gas surface density
is not sensitive to uncertainties resulting from the use of the
$X_{\rm CO}$ conversion factor. These studies in nearby clouds show
that SFR and molecular gas surface density are correlated with a slope
$\sim 2$, which is steeper than that found in extragalactic
studies \citep{Heiderman2010,Gutermuth2011,Lombardi2014}.  A power law
of index $\sim$2 can be explained in terms of thermal fragmentation of
an isothermal self--gravitating layer \citep{Gutermuth2011}.  For gas
surface densities above a certain threshold, the $\Sigma_{\rm
SFR}$--$\Sigma_{\rm gas}$ relationship becomes linear
\citep{Heiderman2010,Lada2010}.

In \citet{Pineda2013} we used [C\,{\sc ii}], H\,{\sc i}, $^{12}$CO,
and $^{13}$CO observations to derive the surface density distribution
of the different phases of the interstellar medium across the Galactic
plane, including warm and cold atomic gas, CO-dark H$_2$ gas, and
CO--traced H$_2$ gas. A study in such detail can currently only be
carried out in the Milky Way. In the following we analyze how the
different phases of the ISM studied in \citet{Pineda2013} are related
to the star formation rate in the Galactic plane with the aim to
connect the results observed in nearby clouds (pc scales) with those
observed in entire galaxies (kpc scales).

  \begin{figure}[t]
%/home/jpineda/got_c+/rot_curves/masks_v2/dark_gas
  \centering
   \includegraphics[width=0.35\textwidth]{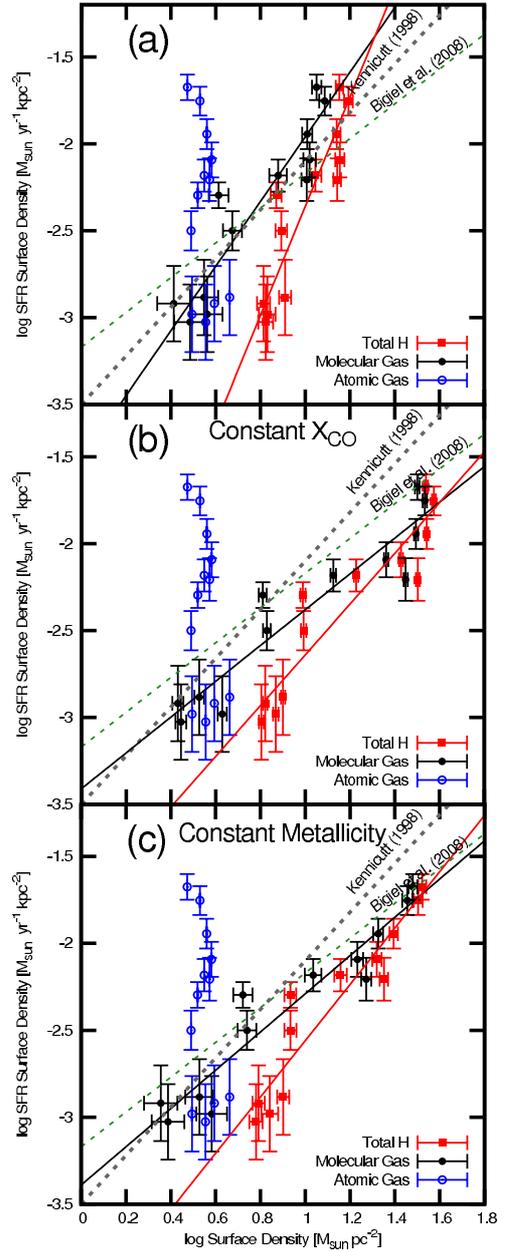}
      \caption{({\it a}) Star formation surface density ($\Sigma_{\rm
      SFR}$) as a function of the molecular, atomic, and total
      hydrogen (H\,{\sc i}+2H$_2$) gas surface density, as determined
      by \citet{Pineda2013} in the Galactic plane. ({\it b})
      $\Sigma_{\rm SFR}$ as function of gas surface density as above,
      but with the molecular and total hydrogen surface densities
      calculated by using the $^{12}$CO $J=1\to0$ intensity
      distribution and a constant CO--to--H$_2$ conversion factor of
      1.74$\times$10$^{20}$ cm$^{-2}$ (K km s$^{-1}$)$^{-1}$. ({\it
      c}) $\Sigma_{\rm SFR}$ as function of gas surface density as
      above, but with $\Sigma_{\rm gas}$ derived assuming constant
      C$^+$ and CO abundances and $^{12}$C/$^{13}$C isotopic ratio.
      The straight lines are linear fits to the data (with matching
      color code). We also include the relationships determined for
      external galaxies by \citet{Kennicutt98} and
      \citet{Bigiel2008}. All gas surface densities have been
      corrected to account for the contribution of He. }
\label{fig:ks_law}
\end{figure}

  \begin{figure}[t]
%/home/jpineda/got_c+/rot_curves/masks_v2/dark_gas
  \centering
   \includegraphics[width=0.48\textwidth]{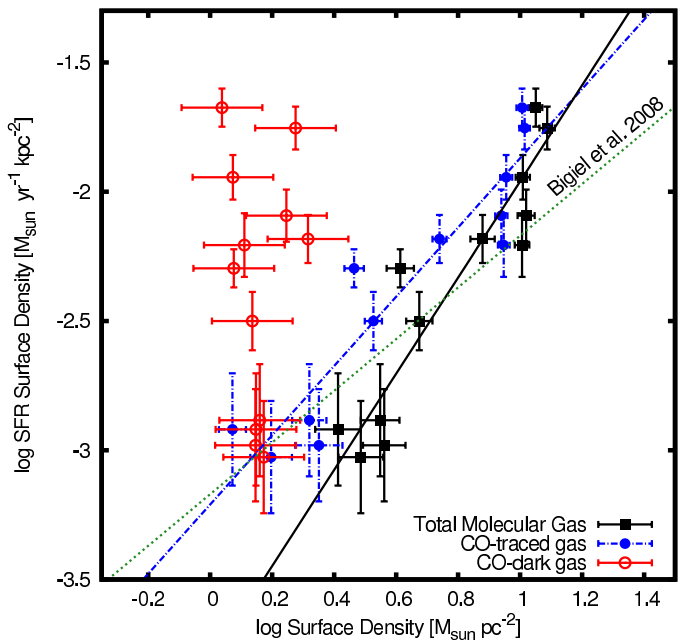}
      \caption{$\Sigma_{\rm SFR}$ as a function of the CO--dark H$_2$,
      CO--traced H$_2$, and total molecular gas surface densities, as
      determined by \citet{Pineda2013}.  The straight lines are linear
      fits to the data. We also include the relationship between the
      SFR and molecular gas surface densities determined for external
      galaxies by \citet{Bigiel2008}. All gas surface densities have
      been corrected to account for the contribution of He. }
\label{fig:ks_law_v2}
\end{figure}

In Figure~\ref{fig:ks_law}a, we show the relationship between the SFR
and the molecular, atomic, and total hydrogen (H\,{\sc i}+2H$_2$)
surface densities as estimated by \citet{Pineda2013}.  We also include
the $\Sigma_{\rm SFR}$--$\Sigma_{\rm gas}$ relationship derived for
external galaxies by \citet{Kennicutt98} (for H\,{\sc i}+2H$_2$) and
by \citet{Bigiel2008} (for H$_2$ only). These relationships, and all
gas surface densities presented here, have been corrected to account
for the contribution of He.  Additionally, for comparison, because the
extragalactic relationships were derived using different values of
$X_{\rm CO}$, we scaled them so they correspond to those derived using
$X_{\rm CO}=1.74\times10^{20}$ cm$^{-2}$\,(K km s$^{-1}$)$^{-1}$
\citep{Grenier2005}. The result of the BES fits are listed in
Table~\ref{tab:sfr_gas}.  Using the $^{12}$CO, $^{13}$CO, and [C\,{\sc
ii}] to derive the molecular gas surface density results in a
relationship with the SFR with a slope $\sim$2 that is steeper than
that derived by \citet{Bigiel2008}, but one that is consistent with
that observed in nearby clouds.  Adding the contribution from H\,{\sc
i} to the surface density results in a slope ($\sim$3.1) that is
steeper than that derived by \citet{Kennicutt98}.  As the gas
depletion time is defined as $\tau^{\rm gas}_{\rm dep}=\Sigma_{\rm
gas}/\Sigma_{\rm SFR}$, the steeper slope of the relationship found
here implies that the gas is converted into stars at a faster rate in
high--$\Sigma_{\rm gas}$ regions.  By itself the atomic gas shows no
correlation with the star formation rate. This lack of correlation
also holds when we separate the atomic gas in the warm and cold
neutral medium components as determined by \citet{Pineda2013}.

In Figure~\ref{fig:ks_law}b, we show the $\Sigma_{\rm SFR} -
\Sigma_{\rm gas}$ relationship resulting when the molecular gas
contribution is estimated by simply multiplying the $^{12}$CO
emissivity distribution by a constant CO--to--H$_2$ conversion factor,
$X_{\rm CO}=1.74\times10^{20}$ cm$^{-2}$\,(K km s$^{-1}$)$^{-1}$, a
method routinely used in extragalactic studies.   We find that the
slope of the $\Sigma_{\rm SFR}-\Sigma_{\rm gas}$ relationship for
molecular gas, $\sim$1.0, is in excellent agreement with that derived
by \citet{Bigiel2008}. Because we use a constant value of $X_{\rm
CO}$, we overestimate the gas surface density as we move inwards in
the Galaxy.  In the case where the contribution from H\,{\sc i} is
considered, we find a relationship with a slope $\sim$1.5 that
approaches the slope derived by \citet{Kennicutt98}.  The differences
in the slope resulting from using different approaches to determine
the molecular gas surface density suggest that the slopes found in
extragalactic studies are strongly influenced by the assumption of a
constant $X_{\rm CO}$ factor.  External galaxies exhibit a wide range
of slopes in the abundance gradients \citep[$-0.1$ to
$-0.9$;][]{Moustakas2010}, with the slope we use for the Milky Way
($-$0.07) being at the lower end of this range. This wide range of
slopes can be an important factor contributing to the scatter in the
$\Sigma_{\rm SFR} - \Sigma_{\rm gas}$ relationship observed in large
samples of external galaxies, when a constant value of $X_{\rm CO}$ is
used.  Note that metallicity is often measured in terms of the oxygen
abundance, while for the Milky Way we use the slope of the abundance
gradient derived by \citet{Rolleston2000} for carbon. The carbon and
oxygen abundance gradients might not be necessarily equal considering
the different production mechanisms of both elements. However, in the
Milky Way, \citet{Rolleston2000} derives an slope for the [O/H]
gradient of $-0.067$ which is similar, within the uncertainties, to
that derived for [C/H] ($-0.07$). Therefore, the results from
\citet{Rolleston2000} suggest that the possible difference between
[O/H] and [C/H] gradients is not significant.  In summary, our results
suggest that the steeper relationships between $\Sigma_{\rm SFR}$ and
the surface density of molecular gas and total hydrogen (H+2H$_2$) gas
found in the Milky Way could also apply to external galaxies.

\begin{table}                                            
\caption{Correlation between the star formation and gas surface
densities for different rings in the Galactic plane obtained using BES fits }
\label{tab:sfr_gas}
\centering						    
\begin{tabular}{l c c c}				    
\hline\hline
Variable  &  \multicolumn{2}{c}{$\log \Sigma_{\rm SFR} = m \log \Sigma_{\rm gas}+b$}     \\
 & $m$  & $b$      \\
\hline
H+2H$_2$                        & 3.1$\pm$0.3 & -5.5$\pm$0.3 \\

H+2H$_2$ Constant $X_{\rm CO}$  &1.5$\pm$0.1 & -4.1$\pm$0.1 \\

H+2H$_2$ Constant Metallicity   &1.6$\pm$0.1 & -4.2$\pm$0.1 \\

\hline

H$_2$                       & 1.9$\pm$0.2 & -3.8$\pm$0.1 \\

H$_2$ Constant $X_{\rm CO}$ & 1.0$\pm$0.1 & -3.4$\pm$0.1 \\

H$_2$ Constant Metallicity  & 1.1$\pm$0.1 & -3.4$\pm$0.1 \\

\hline
H$_2$ CO--traced          & 1.3$\pm$0.1 & -3.2$\pm$0.1\\

\hline

H$_2$ C$^{18}$O--traced dense gas  &  1.0$\pm$0.2 & -3.2$\pm$0.3 \\ 

\hline
\end{tabular}						    
\end{table}

In \citet{Pineda2013}, we showed that $X_{\rm CO}$ varies with
Galactocentric distance (see their Figure 20), and that this variation
is mostly produced by the metallicity gradient of the Galaxy, which
affects the conversion of the $^{12}$CO column density to that of
H$_2$. To study the influence of the metallicity gradient, we
recalculated the molecular gas surface density assuming constant
fractional abundances of [C$^+$]/[H$_2$]$=2.8\times10^{-4}$ and
[CO]/[H$_2$]$=1\times10^{-4}$ (these values are discussed in Section
5.2.2 in \citealt{Pineda2013}) and a constant $^{12}$C/$^{13}$C
isotopic ratio of 65. We compare the SFR and gas surface densities in
this case in Figure~\ref{fig:ks_law}c.  By using a constant abundance,
we find relationships with slopes $\sim$1.1 for H$_2$ gas and
$\sim$1.6 for H+2H$_2$ gas, which are now consistent with those
derived using a constant $X_{\rm CO}$ and therefore with those derived
by \citet{Bigiel2008} and \citet{Kennicutt98}.  Note that the steeper
slope of the $\Sigma_{\rm SFR}$--$\Sigma_{\rm gas}$ relationship in
Figure~\ref{fig:ks_law}a is dominated by the metallicity gradient,
with the $^{12}$C/$^{13}$C ratio gradient playing a minor role.  The
results presented here suggest that metallicity gradients play an
important role setting the slope of the relationship between SFR and
gas surface densities in galaxies.

The variation of the $X_{\rm CO}$ factor as a function of metallicity
in the disk of galaxies has been studied by \citet{Sandstrom2013}, see
also \citet{Magrini2011}, and their results were applied to the
$\Sigma_{\rm SFR}-\Sigma_{\rm gas}$ relationship by
\citet{Leroy2013}. \citet{Sandstrom2013} find no significant
variations in $X_{\rm CO}$ as a function of Galactocentric radius with
the exception of the center of galaxies \citep[see
also][]{Blanc2013}. Note, however, that the gradient in $X_{\rm CO}$
seen in \citet{Pineda2013} suggest a variation of $X_{\rm CO}$ in the
inner Galaxy of about 0.3\,dex which is similar to uncertainties in
the derivation of $X_{\rm CO}$ in \citet{Sandstrom2013}. Thus, if the
magnitude of the $X_{\rm CO}$ variations in the inner disk of galaxies
are similar to those in the Milky Way, then the gradient would have
been obscured by the uncertainties in their derivation of $X_{\rm
CO}$.

In Figure~\ref{fig:ks_law_v2} we show $\Sigma_{\rm SFR}$ as a function
of the CO--dark H$_2$, CO--traced H$_2$ gas, and total molecular gas
surface densities, as determined by \citet{Pineda2013}. The slope
$\sim$2 found for the relationship between star formation and total
molecular gas surface density in Figure~\ref{fig:ks_law}a is the
result of the combination of the contributions from the CO--dark H$_2$
gas and the CO--traced molecular gas. As listed in
Table~\ref{tab:sfr_gas}, the slope of the relationship involving the
CO--traced H$_2$ gas is $\sim$1.3.  Although the CO--dark H$_2$ gas is
not correlated with the star formation rate, its contribution makes
the slope of the relationship between SFR and total molecular gas
steeper, because the CO--dark H$_2$ and CO--traced H$_2$ gas have
similar contributions to the total surface density at low values.

  \begin{figure}[t]
%/home/jpineda/got_c+/rot_curves/masks_v2/dark_gas
  \centering
   \includegraphics[width=0.45\textwidth]{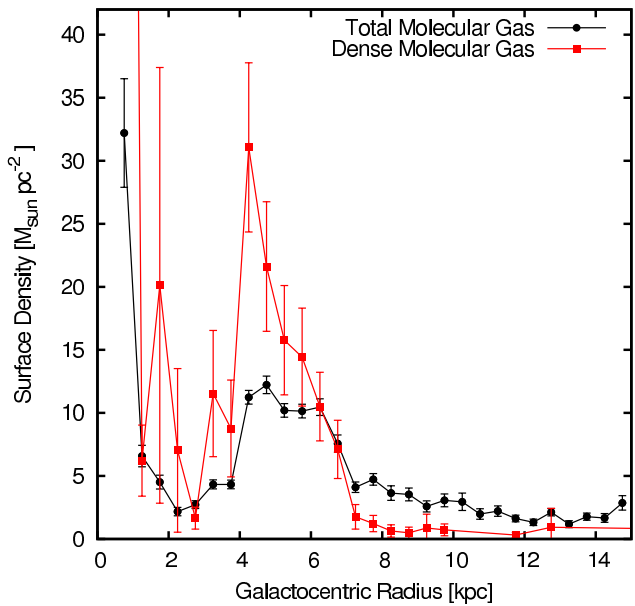}
      \caption{Radial distribution of the azimuthally--averaged
      surface density of dense molecular gas, derived from C$^{18}$O
      observations. We also include the azimuthally--averaged surface
      density distribution of the total H$_2$ gas derived from
      $^{12}$CO, $^{13}$CO, and [C\,{\sc ii}] observations which also
      includes a contribution from diffuse molecular gas.  All gas
      surface densities have been corrected to account for the
      contribution of He. }
\label{fig:surf_dense}
\end{figure}

  \begin{figure}[t]
%/home/jpineda/got_c+/rot_curves/masks_v2/dark_gas
  \centering
   \includegraphics[width=0.45\textwidth]{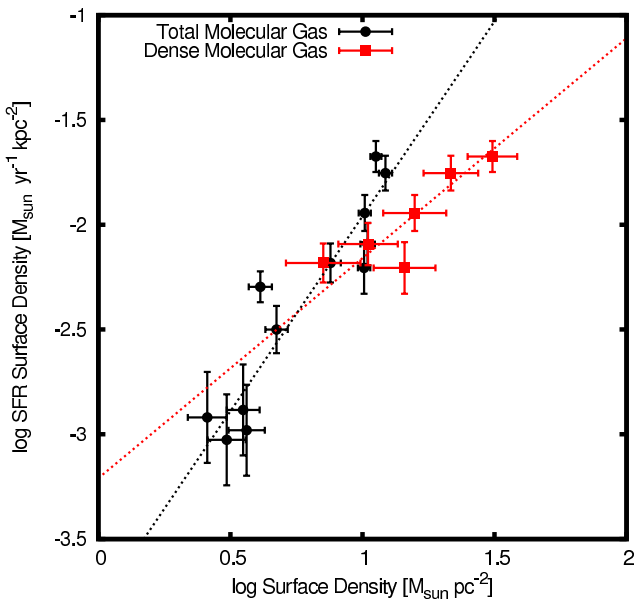}
      \caption{The star formation rate surface density as a function
      of that of dense molecular gas, as estimated from C$^{18}$O
      observations. We also include the $\Sigma_{\rm SFR}-\Sigma_{\rm
      gas}$ relationship for the total molecular gas which also
      accounts for the contribution of diffuse molecular gas.  All gas
      surface densities have been corrected to account for the
      contribution of He.  }
\label{fig:dense_gas_ks}
\end{figure}

%subsection{Dense Molecular Gas}

As mentioned above, the relationship between star formation and gas
surface density in nearby clouds becomes linear above a certain
threshold surface density.  Additionally, observations of dense gas
tracers (e.g. HCN) also suggest a linear relationship between the
dense gas and star formation rate surface density in local clouds
\citep{Wu2005} and external galaxies \citep{Gao2004}.  These results
indicate that there is a different relationship between star formation
and dense molecular gas surface densities compared with that involving
the total molecular gas surface density, which also contains diffuse
molecular gas.  In order to test whether this result also applies to
kpc scales in the Galactic plane, we use observations of C$^{18}$O to
derive the radial distribution of the surface density of dense gas as
shown in Figure~\ref{fig:surf_dense}.

Due to its low abundance, large H$_2$ column densities are needed for
 C$^{18}$O emission to be detected at the sensitivity of our
 observations. Molecular clouds are expected to be compact compared
 with, for example, those of H\,{\sc i} gas. Thus, the large H$_2$
 column densities required to detect C$^{18}$O are likely a result of
 relatively large volume density gas.  The C$^{18}$O observations were
 taken with the Mopra 20m telescope and are described in
 \citet{Pineda2013}. We first calculated the C$^{18}$O column density
 following the procedure described in \citet{Pineda2010a,Pineda2013},
 which first assumes optically thin C$^{18}$O emission, an excitation
 temperature derived from $^{12}$CO where both this line and C$^{18}$O
 are detected, and finally includes an opacity correction. We
 converted the C$^{18}$O column density to a $^{12}$CO column density
 using a $^{12}$CO/C$^{18}$O ratio that varies with Galactocentric
 distance so that it follows the slope derived by \citet{Savage2002}
 for the $^{12}$CO/$^{13}$CO ratio and corresponds to
 $^{12}$CO/C$^{18}$O=557 \citep{Wilson1999} at $R_{\rm
 gal}=8.5$\,kpc. Similarly, we converted the $^{12}$CO column density
 to that of H$_2$ by applying a [CO]/[H$_2$] gradient with a slope
 derived by \citet{Rolleston2000} and [CO]/[H$_2$]=$1\times10^{-4}$ at
 $R_{\rm gal}=8.5$\,kpc.  In Figure~\ref{fig:surf_dense}, we see that
 the azimuthally--averaged distribution of the dense molecular gas is
 similar to that of the total molecular gas derived from $^{12}$CO,
 $^{13}$CO, and [C\,{\sc ii}] (which also includes diffuse molecular
 gas), but that it extends over a narrower range of Galactocentric
 radius (4--7\,kpc) while showing a pronounced peak at about 4.5\,kpc.

In Figure~\ref{fig:dense_gas_ks}, we show how the surface density of
dense H$_2$ gas is related the star formation rate surface density. We
consider rings between 4 and 7\,kpc, which correspond to the range
where most of the dense H$_2$ gas is detected. The result of the BES
fit to the relationship between the dense gas and SFR surface
densities is listed in Table~\ref{tab:sfr_gas}.  Similar to studies of
external galaxies and nearby clouds, the dense gas and SFR surface
densities are almost linearly correlated in the plane of the Milky
Way. We also show in the figure the relationship between star
formation and molecular gas (from $^{12}$CO and $^{13}$CO) to
illustrate the difference in the slopes.

The linear correlation between $\Sigma_{\rm SFR}$ and $\Sigma^{\rm
dense}_{\rm gas}$ can be understood as a consequence of the fact that
massive star clusters form nearly exclusively in dense, massive
cores. The more dense cores a cloud (or galaxy) has, the more massive
stars it will form. The slope of the relationship between $\Sigma_{\rm
SFR}$ and $\Sigma_{\rm gas}$ for the total molecular gas
is steeper than that between $\Sigma_{\rm SFR}$ and $\Sigma^{\rm dense}_{\rm gas}$
because it traces in addition diffuse gas that is not directly
associated with star formation but which dominates the gas surface
density at low values.

\section{Conclusions}
\label{sec:conclusions}

In this paper we studied the relationship between the [C\,{\sc ii}]
  emission and star formation rate in the plane of the Milky Way. We
  also studied how the SFR surface density is related to that of the
  different phases of the ISM. We compared these relationships with
  those observed in external galaxies and local clouds.  Our results
  can be summarized as follows:

\begin{enumerate}

\item We find that the [C\,{\sc ii}] luminosity and SFR are well
correlated on Galactic scales with a relationship that is consistent
with that found for external galaxies.

\item We find that the [C\,{\sc ii}] luminosities arising from the
different phases of the interstellar medium are correlated with the
SFR, but only the combined emission shows a slope that is consistent
with extragalactic observations.

\item We find that the different ISM phases contributing to the total
[C\,{\sc ii}] luminosity of the Galaxy have roughly comparable
contributions: dense PDRs (30\%), cold H\,{\sc i} (25\%), CO--dark
H$_2$ (25\%), and ionized gas (20\%).  The contribution from ionized
gas to the [C\,{\sc ii}] luminosity of the Milky way is larger than
that for the emissivity at $b=0$\degr\ because of the larger scale
height of this ISM component relative to the other contributing
phases.

\item By combining [C\,{\sc ii}] luminosity and SFR data points in the
Galactic plane with those observed in external galaxies and nearby
star forming regions, we find that a single scaling relationship
between the [C\,{\sc ii}] luminosity and SFR,
$\log$(SFR[M$_\odot$\,yr$^{-1}$])=(0.89$\pm$0.04)$\log(L_{\rm
[CII]}$[erg s$^{-1}])-(36.3\pm1.5)$, applies over six orders of
magnitude in these quantities.

\item We find that the [C\,{\sc ii}] cooling rate for the different
contributing ISM components and the star formation rate are well
correlated, as expected for the case of [C\,{\sc ii}] cooling
balancing the gas heating which is dominated by FUV photons from young
stars. The slope in the relationship between cooling rate and SFR is
determined by the radial volume density and metallicity gradients.

\item We studied how star formation is related to the gas surface
 density in the Galactic plane. We find that the SFR and gas surface
 density relationships show a steeper slope compared to that observed
 by \citet{Kennicutt98} and \citet{Bigiel2008}, but one that is
 consistent with that seen in nearby clouds. We find that the
 different slope is a result of the use of a constant CO--to--H$_2$
 conversion factor in the extragalactic studies, which in turn is
 related to the assumption of constant metallicity in galaxies.

\item We find that the star formation rate surface density in the
Milky way is linearly correlated with that of dense molecular gas, as
traced by C$^{18}$O.

\end{enumerate}

The different scaling relationships between star formation, [C\,{\sc
ii}] emission, and gas surface density apply over a large range of
spatial scales going from those of nearby clouds to those of entire
distant galaxies. These multiscale relationships suggest that
understanding the process of star formation locally provides insights
into the process of star formation in the early universe.

\begin{acknowledgements}
 
This research was conducted at the Jet Propulsion Laboratory,
California Institute of Technology under contract with the National
Aeronautics and Space Administration.  We thank the staffs of the ESA
and NASA Herschel Science Centers for their help.  We would like to
thank Roberto Assef, Moshe Elitzur, Tanio Diaz--Santos, Guillermo
Blanc, and Rodrigo Herrera--Camus for enlightening discussions. We
also thank an anonymous referee  for a number of
useful comments. 

\end{acknowledgements}

\bibliographystyle{aa} 
\bibliography{sfr_papers}

\end{document}